\documentclass[aps,prb,reprint,twocolumn,superscriptaddress]{revtex4-2}

\usepackage{graphicx}
\usepackage{amsmath,amssymb,bm}
\usepackage{comment}
\usepackage{url}
\usepackage{hyperref}
\usepackage{comment}
\begin{document}

\title{Deeply nonlinear magnon-photon hybrid excitation}

\author{Dinesh Wagle}
\email{wagled@udel.edu}
\affiliation{Department of Physics and Astronomy, University of Delaware, Newark, DE 19716, USA}
\altaffiliation{Present address: Laboratory for Physical Sciences, 8050 Greenmead Dr., College Park, MD 20740, USA}

\author{Anish Rai}
\affiliation{Department of Physics and Astronomy, University of Delaware, Newark, DE 19716, USA}

\author{M. Benjamin Jungfleisch}
\email{mbj@udel.edu}
\affiliation{Department of Physics and Astronomy, University of Delaware, Newark, DE 19716, USA}

\date{\today}
\begin{abstract}
We investigate the microwave–power dependence of magnon–photon coupling in a yttrium iron garnet-sphere/split-ring-resonator hybrid system at room temperature and demonstrate that nonlinear spin-wave interactions suppress the coupling through power-induced dissipation of magnetostatic modes. At low microwave power, the modes exhibit pronounced level repulsion, evidencing strong coupling to the microwave field. As the power increases, however, magnon linewidth broadening progressively weakens the coupling and ultimately suppresses it entirely below a threshold external magnetic field. We show that this behavior originates from Suhl’s first-order instability: magnetostatic modes, which couple to the resonator, parametrically excites two counter-propagating magnons at half its frequency, causing modes below the threshold external magnetic field to vanish. In contrast, magnon modes above the threshold field remain robust even at high power, as the instability criterion is not satisfied in that regime. These results reveal a well-defined nonlinear boundary for magnon-photon coupled systems and highlight a favorable regime for exploiting nonlinear magnonics for frequency conversion, switching, and other functional magnonic devices.
\end{abstract}


\maketitle

\section{\label{sec:intro} Introduction}

Nonlinear interactions between magnons and other physical platforms---including phonons~\cite{Shklovskij_PhysRevB_2018}, photons~\cite{Lee_PRL_2023}, spin textures~\cite{Wang_PhysRevLett_2021}, and other magnons~\cite{Lendinez2023}---enable the formation and control of novel magnonic states for information storage, manipulation, and processing~\cite{YUAN_PhyReport_2022}. Through these interactions, magnon states can be stabilized, giving rise to exotic phases such as magnon Bose--Einstein condensation~\cite{Nikuni_PRL_2000,Demokritov:2006bx,Bozhko2016} and spin superfluidity~\cite{Sonin_Adv_2010,Yuan_SciAdv_2018}, which support long-range magnon-current transport.

Spin waves are inherently nonlinear, as described by the Landau--Lifshitz--Gilbert (LLG) equation~\cite{Gilbert_IEEETM_2004}. This intrinsic nonlinearity makes them a compelling platform for emerging technologies such as neuromorphic computing~\cite{Papp_NatCom_2021} and frequency conversion via parametric pumping~\cite{Gurevich_1996_Book}. 
Furthermore, nonlinear spin-wave coupling facilitates higher-harmonic generation~\cite{Nikolaev_NatCom_2024,Lendinez2023} and efficient parametric pumping~\cite{Sandweg_2011}. A variety of instability mechanisms can also drive magnon populations to high levels, further enriching the dynamic behavior of nonlinear spin-wave systems~\cite{Anderson:1995dw,Mathieu_PhysRevB_2003,Romero_PhysRevB_2009,Kurebayashi_NatMat_2011}. 

Magnonic nonlinearity arises from the intrinsic nonlinearity of the LLG equation, which gives rise to three- and four-magnon scattering processes~\cite{Rezende_IEEE_1990, Schultheiss_PhysRevLett_2009, Tao_PhysRevB_2025}, collectively known as Suhl instabilities~\cite{Suhl1957}.
Three magnon splitting or confluence processes -- first order Suhl instabilities -- are important in frequency down-conversion and up-conversion, making them essential for applications in magnon-based signal processing and quantum information technologies~\cite{Rao_PhysRevLett_2023, Wang_NatPhys_2024}. This process can also be harnessed to control magnon–microwave photon coupling~\cite{Lee_PRL_2023, Zhang_PhysRevApplied_2023}.

Hybrid quantum systems leverage the complementary strengths of different physical platforms to overcome individual limitations and enable advanced functionalities. Among them, magnon polaritons -- quasiparticles formed through the coupling of magnons with microwave photons~\cite{Artman_PhysRev._1953, Soykal:2010hz, zhang_2014_PRL} -- are particularly significant due to their potential applications in coherent information processing~\cite{Chumak_NatPhys_2015, RAMESHTI_Reports_2022} and the broader field of quantum magnonics~\cite{Lachance_Science_2020, Xu_PhysRevLett_2023, YUAN_PhyReport_2022},  including quantum transduction~\cite{Zhu_optica_2020}, quantum sensors~\cite{zhang_2017_NatComm}, and quantum memories~\cite{BHOI_2019}. For efficient energy exchange between the subsystems of a magnonic hybrid system, strong coupling and high cooperativity are essential~\cite{Awschalom_IEEE_2021}. Additionally, achieving controlled coupling is crucial for the development of on-chip hybrid magnonic devices~\cite{Li_PhysRevLett_2019}.

Magnon polaritons have been widely investigated in the linear-response regime under weak microwave excitation, particularly in the strong and ultrastrong coupling regimes~\cite{Li_JAP_2020, RAMESHTI_Reports_2022}. While suppression of the coupling strength due to nonlinear spin-wave interactions has been observed in yttrium iron garnet (YIG) film/split ring resonator (SRR) systems at elevated microwave powers~\cite{Lee_PRL_2023}, the behavior of magnon polaritons under a \textit{deeply} nonlinear drive  
in other experimental configurations and geometries, including spheres, has remained largely unexplored.

Here, we demonstrate that deeply nonlinear spin-wave interactions suppress magnon–photon coupling in a YIG-sphere/SRR system through nonlinear dissipation of magnetostatic modes above a threshold microwave power {and below a threshold magnetic field}. At low power, the modes are coupled to microwave photons, but as the drive power increases, magnon-linewidth broadening progressively weakens the coupling until it eventually disappears. Once the threshold microwave is exceeded, the modes undergo Suhl’s first-order instability, causing magnon modes below the threshold magnetic field to completely vanish -- only the resonator mode remains detectable. In contrast, modes above the threshold magnetic field remain intact even at high microwave powers, as the criteria for instability are not satisfied in that regime.

\begin{figure}[t]
    \centering
\includegraphics[width=.48\textwidth]{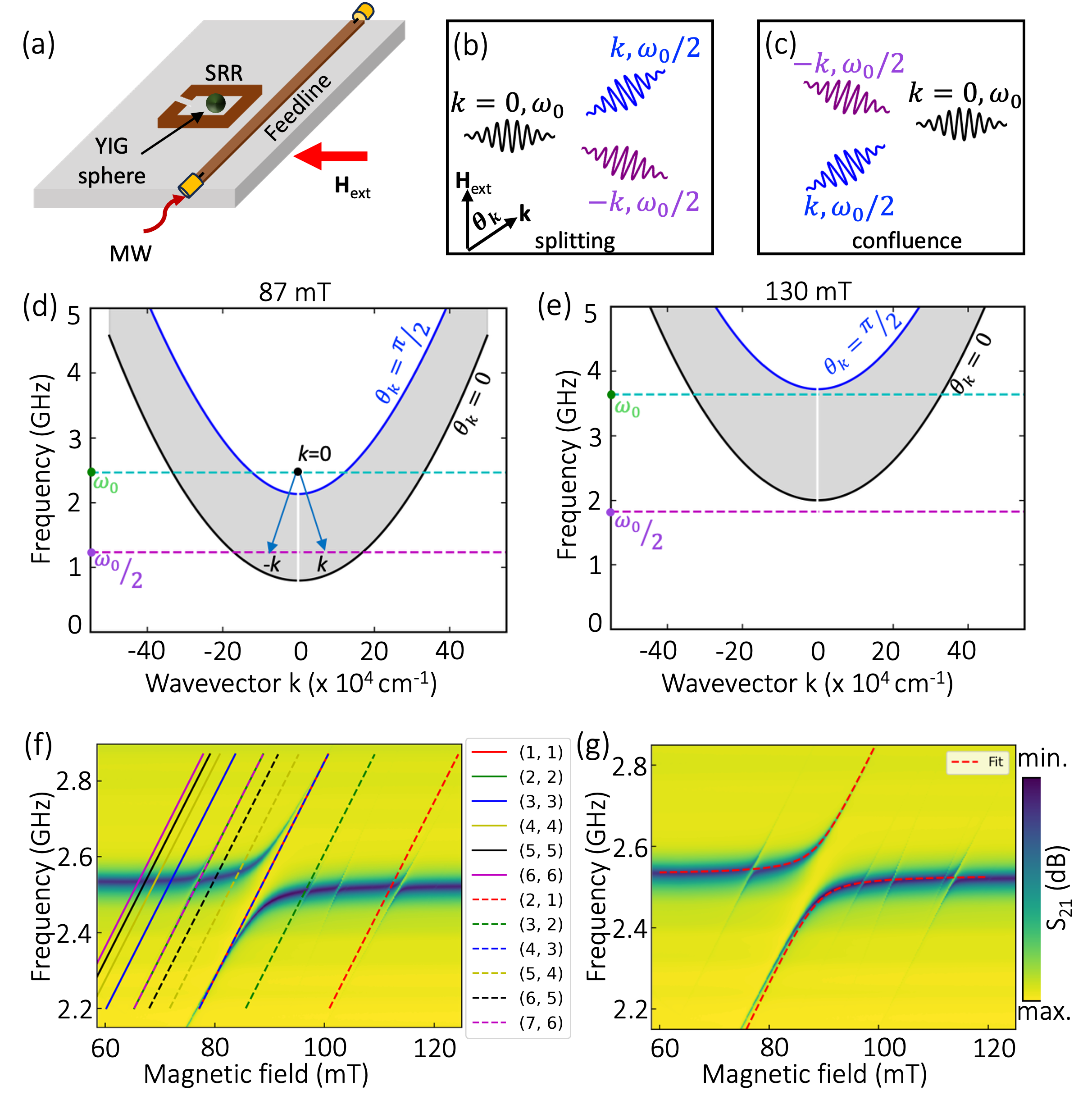}
    \caption{(a) Experimental setup: The split-ring resonator (SRR) is positioned near the microwave feedline, with a YIG sphere placed in the center of the SRR. Transmission spectra are recorded by a VNA as a function of magnetic field and frequency. Schematic illustrations of the  three-magnon (b) splitting and (c) confluence processes in which momentum and energy are conserved.
    YIG sphere spin-wave dispersions at (d) 87 mT and (e) 130 mT, respectively, for propagation angles ranging from $\theta_k = 0$ (black curve) to $\pi/2$ (blue curve). The shaded regions represent intermediate angles. 
    Cyan and magenta dashed lines serve as guides to the eye for the frequencies $\omega_0$ and $\omega_0/2$, respectively. Three magnon splitting is possible at 87 mT, but not at 130 mT. (f) Measured transmission spectra showing the coupling between magnon and the SRR modes at –20 dBm. Overlaid are calculated magnetostatic mode families in the sphere: ($m, m$) modes shown as solid lines and ($m+1, m$) modes as dotted lines. (g) Avoided level crossing corresponding to the (1,1) mode - i.e., Kittel mode - and the SRR mode and a fit to the coupled harmonic oscillator model. 
    }
    \label{fig1}
\end{figure}
\section{Background}
The threshold microwave field for three-magnon scattering is~\cite{Gurevich_1996_Book,JAP_Fletcher_1961}
\begin{equation}\label{threshold field}
    h_{thr} = \min\Bigg[\frac{2\omega_{rk}\omega_p\sqrt{\omega_{r0}^2+(\omega_p-\omega_0)^2}}{\gamma\omega_M\sin 2\theta_k(\omega_p/2+\omega_H+\lambda_{ex} k^2)}\Bigg],
\end{equation}
where $\omega_{r0}$ is the damping rate of the resonantly excited magnetostatic modes (MSM) 
and $\omega_{rk}$  is the damping rate of the spin-wave mode with wavevector $k$, $\omega_p$ is the pumping frequency, and $\omega_0$ is the frequency of the MSM. The parameter $\lambda_{ex}$ represents the exchange stiffness constant. Additionally, $\omega_{M} = -\gamma \mu_0M_s$ and $\omega_H = -\gamma\mu_0H_{ext}$ where $H_{ext}$ is the applied field and $M_s$ is the saturation magnetization, $\gamma$ is the gyromagnetic ratio, and $\theta_k$ is the angle between wavevector ${k}$ and the externally applied field $H_{ext}$.
$h_{thr}$ is minimum when $\omega_p = \omega_0$ and when $\theta_k = 45^\circ$. 

The instability occurs only if~\cite{Gurevich_1996_Book}
\begin{equation}\label{eq:threshold frequency}
    \omega_p < \omega_M\frac{N_z}{2\pi}\equiv \omega_{cr},
\end{equation}
where $N_z$ is the demagnetizing factor in $z$-direction. For a spherical sample, $N_x = N_y = N_z$ and the critical frequency $\omega_{cr} = 2\omega_M/3$, which corresponds to approximately 3.27 GHz for a YIG sphere with $\mu_0 M_s$ = 178 mT. Furthermore, for the first order Suhl instability to occur, the frequency $\omega_p/2$ must lie above the lower boundary of the spin-wave spectrum. In the case of an ellipsoidal sample, this requirement is met when the external magnetic field $H_{ext}$ satisfies the relation~\cite{Gurevich_1996_Book}
\begin{equation}\label{threshold field}
    H_{ext} < \frac{\omega_p}{2\gamma} + N_zM_s.
\end{equation}

\begin{figure*}[t]
    \centering
    \includegraphics[width=1.6\columnwidth]{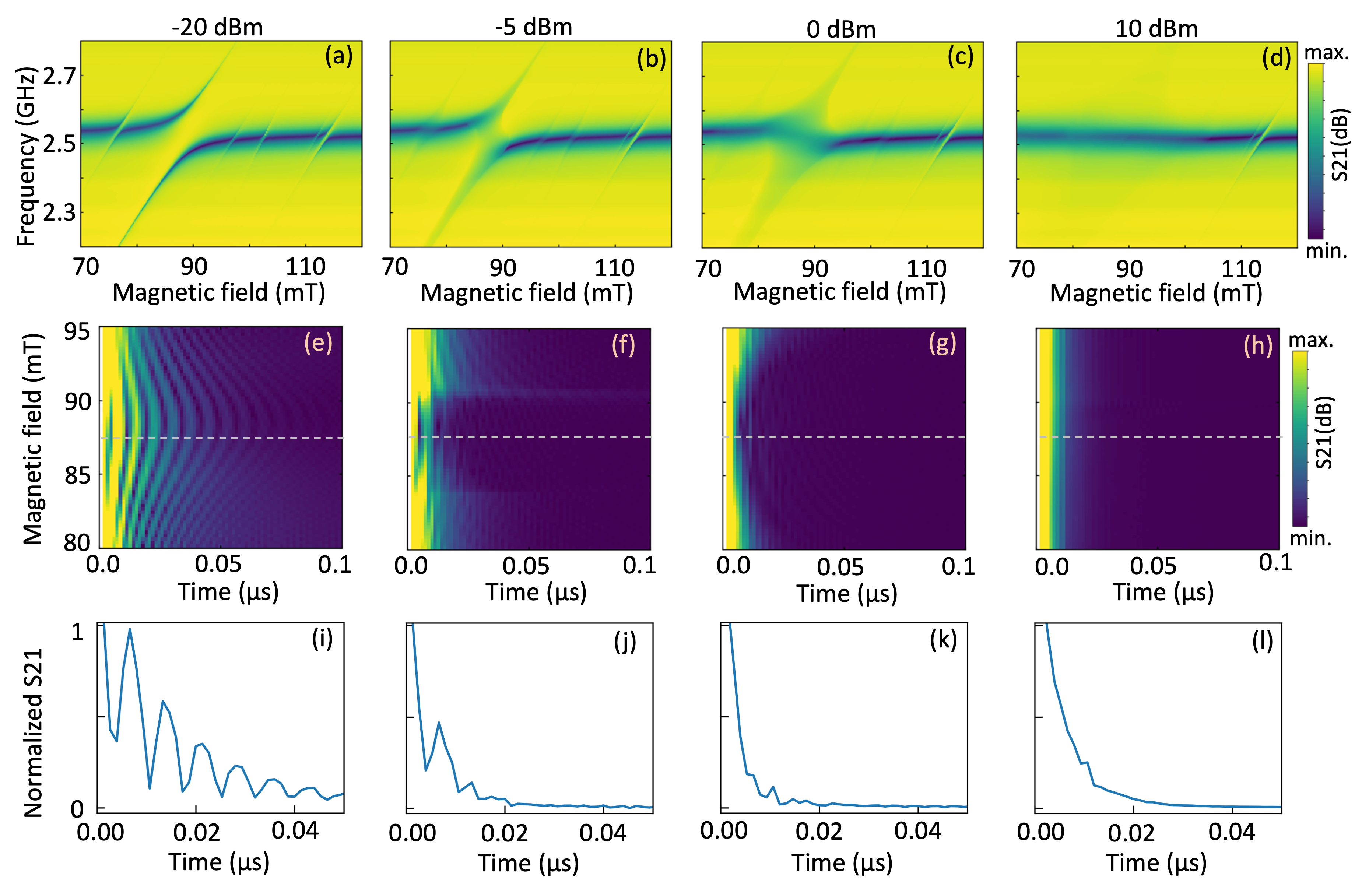}
    \caption{Experimentally observed microwave-power dependent magnon-photon coupling. (a)-(d) Transmission spectra as a function of magnetic field and microwave frequency at input microwave powers of –20, –5, 0, and +10 dBm, respectively. (e)-(h) Corresponding inverse Fourier transforms of panels (a)-(d). (i)-(l) Rabi-like oscillations at 87 mT extracted from the time-domain data in panels (e–h) -- extracted along the dashed lines.}
    \label{fig2}
\end{figure*}

\section{Experiment}\label{experiment}
The magnonic hybrid system comprises a SRR loaded with an epitaxial YIG sphere, see Fig.~\ref{fig1}(a). The SRR was fabricated by etching one side of a Rogers RO3010 laminate with a dielectric constant of 10.20$\pm$0.30 and copper thickness of 17.5 $\mu$m coated on both sides of the substrate. The SRR with a quality factor of 70 was designed to exhibit a resonance frequency of $\sim$2.5 GHz. This design supports both three-magnon splitting and confluence [Figs.~\ref{fig1}(b) and (c)], and ensures that the minimum of the spin-wave dispersion curve lies below half the SRR resonance frequency at the target external magnetic field (e.g., 87 mT)—satisfying the threshold condition given in Eq.~(\ref{eq:threshold frequency}). 
  
The spin-wave dispersions of a YIG sphere at external fields of $H_{\mathrm{ext}}$ = 87 mT and 130 mT, respectively, for spin waves propagating at angles $\theta_k$ ranging from 0 (black solid line) to $\pi/2$ (blue solid line) are shown in Figs.~\ref{fig1}(d) and (e). Interestingly, at lower magnetic fields (e.g., 87 mT), the uniform magnon mode frequency $\omega_0$ lies outside the spin-wave dispersion manifold, with $\omega_0/2$ positioned above the minimum frequency of the dispersion. In contrast, at higher magnetic fields (e.g., 130 mT), $\omega_0$ falls within the dispersion manifold, and $\omega_0/2$ lies below its minimum~\cite{Gurevich_1996_Book}. While this behavior is often illustrated using a continuous spin-wave dispersion relation, it is important to note that in a spherical geometry, the spin-wave modes are inherently discrete. Each of these MSMs is characterized by a specific set of mode indices 
($n, m$), and its frequency is obtained from a characteristic equation {in terms of associated Legendre function $P_n^m$; here, $n$ and $m$ are the spherical polar and azimuthal indices, respectively}. (1,1) corresponds to the special case of the uniform precession, or Kittel mode.

A commercially available single crystal YIG sphere of diameter 0.75 mm was positioned at the center of the SRR to ensure maximum interaction between the magnon and photon modes. An external static magnetic field was applied perpendicular to the SRR feedline. A microwave current, generated by a vector network analyzer (VNA), was applied to one end of the SRR feedline and the resulting transmission was measured at the opposite end by recording the S21 transmission parameter. To explore the effect of high microwave power on the magnon-photon coupling, the microwave power was systematically varied from nominally $-20$~dBm to $+13$~dBm. At each power level, the transmission spectra were recorded to capture the changes in the coupling dynamics.

\section{Results}
Figures~\ref{fig1}(f) and (g) present the measured transmission spectra as a function of magnetic field and microwave frequency at an input power of –20 dBm. The color scale corresponds to the S${21}$ transmission parameter. Avoided crossings in the spectra reveal the coupling between YIG sphere MSMs and the SRR mode. Alongside the dominant avoided crossing, several weaker avoided crossings appear due to a coupling of the resonator to higher-order MSMs. 
These higher-order modes are excited as a result of the nonuniform microwave magnetic field across the sample~\cite{Wagle_JPM_2024}. The spacing between the modes decreases at lower magnetic fields and increases at higher fields. 

In general, MSM resonant frequencies depend nonlinearly on the external magnetic field, except for the mode families with $n-|m|$ = 0 and 1; accordingly, to identify the observed magnon modes, we focus on the MSM families~\cite{Walker_JAP_1958, Fletcher_JAP_1959, Fletcher_PhysRev_1959, Leo_PRB_2020}:
\begin{equation}
    \frac{f}{\gamma M_s}-\frac{H_{ext,mm}}{M_s}+\frac{1}{3}=\frac{m}{2m+1};\,\,\,(n=m), 
    \label{eq:MSM n=m}
\end{equation}
\begin{equation}
    \frac{f}{\gamma M_s}-\frac{H_{ext,m(m+1)}}{M_s}+\frac{1}{3}=\frac{m}{2m+3};\,\,\,(n=m+1).
    \label{eq:MSM n=m+1}
\end{equation}

The MSMs are overlaid on the measured spectra in Fig.~\ref{fig1}(f) using Eqs.~(\ref{eq:MSM n=m}) and (\ref{eq:MSM n=m+1}). The ($m, m$) modes are plotted as solid lines, while the ($m+1, m$) modes are shown as dotted lines. The most prominent avoided crossing arises from the coupling between the SRR mode and the (1,1) uniform mode, highlighted by the red solid line. The ($m, m$) family is degenerate with the ($3m+1, 3m$) family; therefore, the (1,1) uniform mode is degenerate with the (4,3) MSM.

The red dashed curve in Fig.~\ref{fig1}(g) represents the fitted result based on the coupled harmonic oscillator model~\cite{Kaffash_QST_2023}, applied to the avoided crossing associated with the (1,1) uniform mode. This fit accurately captures the characteristic avoided level crossing in the transmission spectra near the resonance. From the fit, a coupling strength $g$ of 70 MHz is obtained, which quantifies the interaction between the uniform precession mode of the YIG sphere and the SRR mode. Since $g$ exceeds both the resonator dissipation rate $\kappa_p$ = 38 MHz and the magnon dissipation rate $\kappa_m$ = 6 MHz (fittings not shown here), the system exhibits a cooperativity $C = g^2/(\kappa_p \kappa_m) > 1$, confirming that the system operates in the strong coupling regime.

\begin{figure}[t]
    \centering
    \includegraphics[width=.48\textwidth]{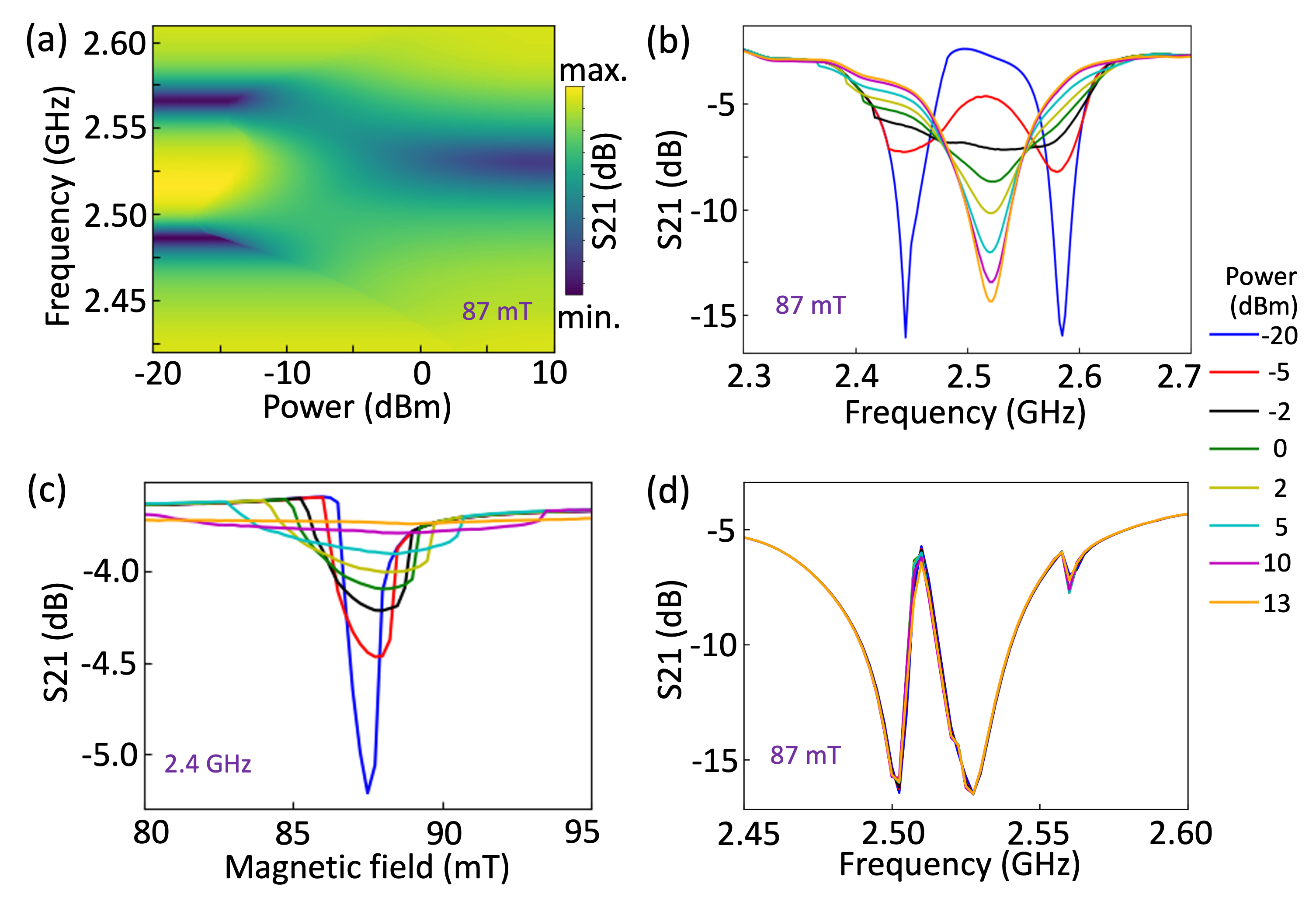}
    \caption{ (a) The transmission spectra as a function of frequency and microwave power at 114 mT. (b) Lineplots of transmission spectra at 87 mT for various microwave power levels extracted from Fig.~\ref{fig2}, showing the evolution of resonance features. (c) Transmission spectra lineplots of the magnon mode detuned from the coupling region as a function of external magnetic field at a fixed frequency of 2.4 GHz, presented for different microwave power levels. (d) Transmission spectra as a function of frequency at 114 mT for various microwave power levels, showing no noticeable power dependence in the spectral response—indicating that Suhl’s first-order instability condition is not satisfied. The small kink above 2.55 GHz is from another nearby higher order mode. }
    \label{fig:M-P_nonlinear2}
\end{figure}

The comparative study of the transmission spectra at different microwave power levels of $-20$ dBm, $-5$ dBm, 0 dBm, and +10 dBm are presented in Figs.~\ref{fig2} (a)-(d). At a nominal power of $-20$ dBm [identical results as in Fig.~\ref{fig1} (f) and (g)], a distinct repulsion gap (coupling strength of 70 MHz) is observed around 87 mT, signifying coherent coupling between the YIG uniform mode and the SRR mode, as discussed in the previous paragraph. However, as the input microwave power is increased to $–5$ dBm [Fig.~\ref{fig2}(b)], the linewidth of the magnon mode broadens noticeably and the avoided crossing gap begins to narrow. These effects lead to a gradual smearing of the spectral features, eventually forming a \textit{bridge-like} structure that connects the upper and lower branches of the avoided crossing. This feature becomes more pronounced at 0 dBm [Fig.~\ref{fig2}(c)], where the magnon broadening increases significantly, and the gap closes. At +10 dBm [Fig.~\ref{fig2}(d)], the SRR mode flattens out, and the magnon modes vanish below $\sim$105 mT, making the spectrum resemble that of a standalone resonator in this field range. Interestingly, a smaller avoided crossing at $\sim$114 mT remains unaffected by the increase in microwave power, which will be discussed later.

The corresponding inverse Fourier transforms of the spectra presented in Figs.~\ref{fig2}(a)-(d) in the magnetic field range of 80 mT to 95 mT are shown in Figs.~\ref{fig2}(e)-(h). At $-20$~dBm [Fig.~\ref{fig2}(e)], a pronounced buildup of magnon amplitude is observed. The time-dependent variation of S${21}$ at 87 mT, taken along the dashed line in Fig.~\ref{fig2}(e), is shown in Fig.~\ref{fig2}(i). Clear Rabi-like oscillations are evident, accompanied by an overall exponential decay, indicating both coherent energy exchange and dissipation in the system. As the microwave power increases, the temporal evolution of the magnon dynamics is progressively altered, see Figs.~\ref{fig2}(f)-(h). The Rabi-like oscillations, Figs.~\ref{fig2}(j)-(l), become increasingly suppressed, indicating a weakening of the coherence between the magnon and SRR modes. At +10 dBm, the time-domain trace reflects an essentially uncoupled SRR response: the Rabi-like oscillations are fully suppressed, and only the exponential decay associated with the isolated SRR mode remains.

We further investigated the power-dependent behavior in the coupling region using smaller power increments to capture the evolution of the spectral features more precisely. The transmission parameter as a function of microwave power and frequency in the coupling region at a fixed magnetic field of 87 mT is shown in Fig.~\ref{fig:M-P_nonlinear2}(a). The corresponding line plots extracted from the transmission spectra at various microwave power levels and for a fixed magnetic field of 87 mT are shown in Fig.~\ref{fig:M-P_nonlinear2}(b). At lower power levels, two strongly coupled modes are observed. As the microwave power increases beyond $\sim -10$ dBm, both coupled modes begin to broaden significantly while approaching one another. This behavior points to power- induced mechanisms -- including enhanced dissipation and nonlinear interactions -- governing the system’s response. At 0 dBm, the magnon mode is largely suppressed, and the transmission spectrum reflects only the SRR mode. Beyond the minimum of $\sim-2$ dBm, the transmission parameter of the SRR mode increases with increasing power. Our observations for a YIG sphere closely resemble that observed in an YIG thin film~\cite{Lee_PRL_2023}. However, under strongly nonlinear driving conditions, the magnetostatic modes in our study completely disappear once the microwave power exceeds a critical threshold.

Next, we will discuss the magnon-mode behavior in a region slightly detuned from the coupling point. Figure~\ref{fig:M-P_nonlinear2}(c) presents the transmission parameter versus applied magnetic field at a fixed microwave frequency of 2.4 GHz for different input powers. Since the resonator frequency lies at 2.5 GHz, we excite and detect only magnon modes in this measurement. The plots clearly reveal how the magnon modes evolve with increasing microwave power. At low microwave power, the magnon mode is both intense and sharply defined. As the microwave power increases, the magnon mode broadens and the transmission drops significantly. At the highest power level of +13 dBm, the magnon mode disappears entirely from the spectrum. Additionally, a foldover effect emerges in the magnon mode as the power increases. This manifests as a shift of the magnon resonance toward higher magnetic fields, accompanied by an increasing asymmetry in the lineshape at elevated power levels. Such foldover behavior and the associated frequency shifts are signatures of nonlinear magnetization dynamics, arising when the magnetization precession enters the nonlinear regime under strong microwave excitation~\cite{Wang_2018_PRL}.

\section{Discussions}
At low microwave powers, the uniform magnon mode (1,1) and higher-order MSMs ($n,m$) in the YIG sphere are coherently excited and strongly couple to the SRR mode, giving rise to well-defined avoided crossings in the transmission spectra. However, as the microwave power is increased, nonlinear effects begin to dominate. Notably, the linewidth of the uniform mode broadens significantly, indicating increased damping and energy loss. This is a hallmark of three-magnon scattering processes, specifically the first order Suhl instability, wherein a MSM mode decays into two magnons with opposite wavevectors $\pm k$, each carrying half the original frequency [see Fig.~\ref{fig1}(b) and (d)] \cite{JAP_Fletcher_1961}. This decay pathway effectively drains energy from the coherent uniform mode into finite $k$ magnons, making the uniform mode appear lossy and weakening its coupling to the SRR.

A similar power-dependent suppression is observed for other MSMs located at bias fields below $\sim$104 mT. These modes are also susceptible to the same three-magnon decay mechanism, as their frequencies and wavevector configurations satisfy the conditions for Suhl's instability, see Eqs.~(\ref{eq:threshold frequency}) and (\ref{threshold field}). We note that the high-field MSMs exhibit a higher $h_{thr}$ than the low-field modes, consistent with previous reports in the literature \cite{JAP_Fletcher_1961}.
In contrast, the (2, 1)-mode, which appears at a bias field of $\sim$114 mT, does not exhibit this behavior, see Fig.~\ref{fig2}. This observation can be understood by considering the threshold condition given by Eq.~(\ref{threshold field}), giving a critical field value of $\sim$ 104 mT in our case. Above this threshold magnetic field, $\omega_0/2$ lies below the bottom of the magnon dispersion, and decay into $\pm k$ modes is no longer possible, as shown in Fig.~\ref{fig1}(e) for 130 mT. Therefore, the (2, 1)-mode at 114 mT remains unaffected by high microwave powers. The (2, 1)-mode continues to exhibit coherent coupling to the SRR mode, as observed in Fig.~\ref{fig:M-P_nonlinear2}(d) for all microwave powers. In contrast to Fig.~\ref{fig:M-P_nonlinear2}(b), where magnon modes near a bias field of $\sim$87 mT undergo damping due to three-magnon scattering, the intensity and linewidth of the coupled modes at 114 mT remain largely unchanged with increasing microwave power, indicating the absence of significant nonlinear interactions. Unfortunately, due to the limitations of the current experimental setup, it is not possible to directly observe the magnons generated as a result of three-magnon scattering at $\pm k$ wavevectors. However, recent demonstrations of pump-induced coupling between the {MSMs} and a magnon condensate -- facilitated by coherent three-magnon splitting and confluence processes -- offer a promising pathway for directly detecting such nonlinear magnons~\cite{Tao_PhysRevB_2025}. Another approach is to detect non-zero wavevector magnons optically by wavevector-resolved Brillouin light scattering spectroscopy~\cite{Serga_PRB}. 

\section{Conclusion}
In summary, we investigated magnon–photon coupling in the nonlinear regime, with an emphasis on how high microwave power alters the coupling dynamics.  
At low powers, two well-resolved hybrid modes -- indicative of strong coupling -- are clearly visible. As the microwave power increases, the magnon mode broadens substantially and, at sufficiently high power, becomes strongly damped and ultimately vanishes, leaving only the resonator mode in the transmission spectrum.

We attribute this power-dependent evolution to Suhl’s first-order instability, a nonlinear three-magnon scattering process activated above a critical threshold of microwave power and magnetic field. In this process, a MSM decays into two magnons of opposite wavevector ($\pm k$) and half the original frequency, conserving both energy and momentum. The pronounced linewidth broadening observed in our experiments is a definitive signature of this instability, confirming the emergence of nonlinear magnon dynamics in the hybrid system.

\section*{Data availability}
The data that support the findings of this article are openly available at~{\url{https://doi.org/10.17605/OSF.IO/QGXFW}}.

\section*{acknowledgments}
We acknowledge support by the U.S. Department of Energy, Office of Basic Energy Sciences, Division of Materials Sciences and Engineering under Award DE-SC0020308. ChatGPT 5.2 (OpenAI) was used to assist with language editing and text refinement. No scientific content or conclusions were generated by the AI tool.

\bibliographystyle{apsrev4-2}
\bibliography{biblio}

@PREAMBLE{
 "\providecommand{\noopsort}[1]{}" 
 # "\providecommand{\singleletter}[1]{#1}%" 
}

@article{JAP_Fletcher_1961,
    author = {Fletcher, P. C. and Silence, Neal},
    title = {Subsidiary Absorption above Ferrimagnetic Resonance},
    journal = {Journal of Applied Physics},
    volume = {32},
    number = {4},
    pages = {706-711},
    year = {1961},
    month = {04},
    issn = {0021-8979},
    doi = {10.1063/1.1736075},
    url = {https://doi.org/10.1063/1.1736075},
 }

@article{Serga_PRB,
  title = {Brillouin light scattering spectroscopy of parametrically excited dipole-exchange magnons},
  author = {Serga, A. A. and Sandweg, C. W. and Vasyuchka, V. I. and Jungfleisch, M. B. and Hillebrands, B. and Kreisel, A. and Kopietz, P. and Kostylev, M. P.},
  journal = {Phys. Rev. B},
  volume = {86},
  issue = {13},
  pages = {134403},
  numpages = {8},
  year = {2012},
  month = {Oct},
  publisher = {American Physical Society},
  doi = {10.1103/PhysRevB.86.134403},
  url = {https://link.aps.org/doi/10.1103/PhysRevB.86.134403}
}

@Article{Lendinez2023,
author={Lendinez, Sergi
and Kaffash, Mojtaba T.
and Heinonen, Olle G.
and Gliga, Sebastian
and Iacocca, Ezio
and Jungfleisch, M. Benjamin},
title={Nonlinear multi-magnon scattering in artificial spin ice},
journal={Nature Communications},
year={2023},
month={Jun},
day={09},
volume={14},
number={1},
pages={3419},
issn={2041-1723},
doi={10.1038/s41467-023-38992-7},
url={https://doi.org/10.1038/s41467-023-38992-7}
}

@Article{Bozhko2016,
author={Bozhko, Dmytro A.
and Serga, Alexander A.
and Clausen, Peter
and Vasyuchka, Vitaliy I.
and Heussner, Frank
and Melkov, Gennadii A.
and Pomyalov, Anna
and L'vov, Victor S.
and Hillebrands, Burkard},
title={Supercurrent in a room-temperature Bose--Einstein magnon condensate},
journal={Nature Physics},
year={2016},
month={Nov},
day={01},
volume={12},
number={11},
pages={1057-1062},
issn={1745-2481},
doi={10.1038/nphys3838},
url={https://doi.org/10.1038/nphys3838}
}

@article{Sandweg_2011,
  title = {Spin Pumping by Parametrically Excited Exchange Magnons},
  author = {Sandweg, C. W. and Kajiwara, Y. and Chumak, A. V. and Serga, A. A. and Vasyuchka, V. I. and Jungfleisch, M. B. and Saitoh, E. and Hillebrands, B.},
  journal = {Phys. Rev. Lett.},
  volume = {106},
  issue = {21},
  pages = {216601},
  numpages = {4},
  year = {2011},
  month = {May},
  publisher = {American Physical Society},
  doi = {10.1103/PhysRevLett.106.216601},
  url = {https://link.aps.org/doi/10.1103/PhysRevLett.106.216601}
}

@article{Papp_NatCom_2021,
author = {Ádám Papp and Wolfgang Porod and Gyorgy Csaba},
title = {Nanoscale neural network using non-linear spin-wave interference},
journal = {Nat. Commun.},
volume = {12},
pages = {6422},
year = {2021}
}

@article{Wagle_JPM_2024,
year = {2024},
month = {feb},
publisher = {IOP Publishing},
volume = {7},
number = {2},
pages = {025005},
author = {Dinesh Wagle and Anish Rai and Mojtaba T Kaffash and M Benjamin Jungfleisch},
title = {Controlling magnon-photon coupling in a planar geometry},
journal = {Journal of Physics: Materials},
}

@book{Gurevich_1996_Book,
  title={Magnetization oscillations and waves},
  author={Gurevich, Alexander G and Melkov, Gennadii A},
  year={1996},
  publisher={CRC press}
}

@article{Walker_JAP_1958,
  title={Resonant modes of ferromagnetic spheroids},
  author={L. R. Walker},
  journal={J. Appl. Phys.},
  volume={2},
  number={},
  pages={318},
  year={1958},
  publisher={}
}

@article{Fletcher_JAP_1959,
    author = {Fletcher, P. C. and Bell, R. O.},
    title = {Ferrimagnetic Resonance Modes in Spheres},
    journal = {Journal of Applied Physics},
    volume = {30},
    number = {5},
    pages = {687-698},
    year = {1959},
    month = {05},
    issn = {0021-8979},
    doi = {10.1063/1.1735216}
}

@article{Leo_PRB_2020,
  title = {Identification and time-resolved study of ferrimagnetic spin-wave modes in a microwave cavity in the strong-coupling regime},
  author = {Leo, Angelo and Monteduro, Anna Grazia and Rizzato, Silvia and Martina, Luigi and Maruccio, Giuseppe},
  journal = {Phys. Rev. B},
  volume = {101},
  issue = {1},
  pages = {014439},
  numpages = {8},
  year = {2020},
  month = {Jan},
  publisher = {American Physical Society},
}

@article{Lee_PRL_2023,
  title = {Nonlinear Magnon Polaritons},
  author = {Lee, Oscar and Yamamoto, Kei and Umeda, Maki and Zollitsch, Christoph W. and Elyasi, Mehrdad and Kikkawa, Takashi and Saitoh, Eiji and Bauer, Gerrit E. W. and Kurebayashi, Hidekazu},
  journal = {Phys. Rev. Lett.},
  volume = {130},
  issue = {4},
  pages = {046703},
  numpages = {6},
  year = {2023},
  month = {Jan},
  publisher = {American Physical Society},
  doi = {10.1103/PhysRevLett.130.046703}
}

@article{Wang_2018_PRL,
  title={Bistability of cavity magnon polaritons},
  author={Wang, Yi-Pu and Zhang, Guo-Qiang and Zhang, Dengke and Li, Tie-Fu and Hu, C-M and You, JQ},
  journal={Physical review letters},
  volume={120},
  number={5},
  pages={057202},
  year={2018},
  publisher={APS}
}

@article{Tao_PhysRevB_2025,
  title = {Pump-induced magnon anticrossing due to three-magnon splitting and confluence},
  author = {Qu, Tao and Xiong, Yuzan and Zhang, Xufeng and Li, Yi and Zhang, Wei},
  journal = {Phys. Rev. B},
  volume = {111},
  issue = {18},
  pages = {L180410},
  numpages = {7},
  year = {2025},
  month = {May},
  publisher = {American Physical Society},
  doi = {10.1103/PhysRevB.111.L180410}
}

@article{Artman_PhysRev._1953,
  title = {Measurement of Permeability Tensor in Ferrites},
  author = {Artman, J. O. and Tannenwald, P. E.},
  journal = {Phys. Rev.},
  volume = {91},
  issue = {4},
  pages = {1014--1015},
  numpages = {0},
  year = {1953},
  month = {Aug},
  publisher = {American Physical Society}
}

@article{Soykal:2010hz,
author = {Soykal, {\"O} O and Flatt{\'e}, M E},
title = {{Strong Field Interactions between a Nanomagnet and a Photonic Cavity}},
journal = {Physical Review Letters},
year = {2010},
volume = {104},
number = {7},
pages = {077202},
month = feb,
publisher = {American Physical Society},
doi = {10.1103/PhysRevLett.104.077202}
}

@article{zhang_2014_PRL,
  title={Strongly coupled magnons and cavity microwave photons},
  author={Zhang, Xufeng and Zou, Chang-Ling and Jiang, Liang and Tang, Hong X},
  journal={Physical review letters},
  volume={113},
  number={15},
  pages={156401},
  year={2014},
  publisher={APS}
}

@article{zhang_2017_NatComm,
  title={Observation of the exceptional point in cavity magnon-polaritons},
  author={Zhang, Dengke and Luo, Xiao-Qing and Wang, Yi-Pu and Li, Tie-Fu and You, JQ},
  journal={Nature communications},
  volume={8},
  number={1},
  pages={1368},
  year={2017},
  publisher={Nature Publishing Group UK London}
}

@article{Zhu_optica_2020,
author = {Na Zhu and Xufeng Zhang and Xu Han and Chang-Ling Zou and Changchun Zhong and Chiao-Hsuan Wang and Liang Jiang and Hong X. Tang},
journal = {Optica},
number = {10},
pages = {1291--1297},
publisher = {Optica Publishing Group},
title = {Waveguide cavity optomagnonics for microwave-to-optics conversion},
volume = {7},
month = {Oct},
year = {2020},
}

@article{Lachance_Science_2020,
  title={Entanglement-based single-shot detection of a single magnon with a superconducting qubit},
  author={Lachance-Quirion, Dany and Wolski, Samuel Piotr and Tabuchi, Yutaka and Kono, Shingo and Usami, Koji and Nakamura, Yasunobu},
  journal={Science},
  volume={367},
  number={6476},
  pages={425--428},
  year={2020},
  publisher={American Association for the Advancement of Science}
}

@article{Xu_PhysRevLett_2023,
  title = {Quantum Control of a Single Magnon in a Macroscopic Spin System},
  author = {Xu, Da and Gu, Xu-Ke and Li, He-Kang and Weng, Yuan-Chao and Wang, Yi-Pu and Li, Jie and Wang, H. and Zhu, Shi-Yao and You, J. Q.},
  journal = {Phys. Rev. Lett.},
  volume = {130},
  issue = {19},
  pages = {193603},
  numpages = {6},
  year = {2023},
  month = {May},
  publisher = {American Physical Society},
  doi = {10.1103/PhysRevLett.130.193603},
  url = {https://link.aps.org/doi/10.1103/PhysRevLett.130.193603}
}

@article{YUAN_PhyReport_2022,
title = {Quantum magnonics: When magnon spintronics meets quantum information science},
author = {H.Y. Yuan and Yunshan Cao and Akashdeep Kamra and Rembert A. Duine and Peng Yan},
journal = {Physics Reports},
volume = {965},
pages = {1-74},
year = {2022},
issn = {0370-1573},
doi = {https://doi.org/10.1016/j.physrep.2022.03.002},
url = {https://www.sciencedirect.com/science/article/pii/S0370157322000977},
}

@incollection{BHOI_2019,
title = {Chapter One - {P}hoton-magnon coupling: Historical perspective, status, and future directions},
series = {Solid State Physics},
publisher = {Academic Press},
volume = {70},
pages = {1-77},
year = {2019},
booktitle = {Recent Advances in Topological Ferroics and their Dynamics},
author = {Biswanath Bhoi and Sang-Koog Kim}
}

@ARTICLE{Awschalom_IEEE_2021,
  author={Awschalom, David D. and Du, Chunhui Rita and others},
  journal={IEEE Transactions on Quantum Engineering}, 
  title={Quantum Engineering With Hybrid Magnonic Systems and Materials (Invited Paper)}, 
  year={2021},
  volume={2},
  number={},
  pages={1-36},
  }

@article{Chumak_NatPhys_2015,
author = {Chumak, A V and Vasyuchka, V I and Serga, A A and Hillebrands, B},
title = {{Magnon spintronics}},
journal = {Nature Physics},
year = {2015},
volume = {11},
number = {6},
pages = {453--461}
}

@article{Gilbert_IEEETM_2004,
author = {Gilbert, T L},
title = {{A phenomenological theory of damping in ferromagnetic materials}},
journal = {IEEE Trans. Magn.},
year = {2004},
volume = {40},
pages = {3443},
uri = {\url{papers3://publication/uuid/DBCA5A92-DB3A-4A8A-A680-7B039BC1FE14}}
}

@article{Nikuni_PRL_2000,
  title = {{Bose-Einstein Condensation of Dilute Magnons in $\mathrm{TlCuCl}_{3}$}},
  author = {Nikuni, T. and Oshikawa, M. and Oosawa, A. and Tanaka, H.},
  journal = {Phys. Rev. Lett.},
  volume = {84},
  issue = {25},
  pages = {5868--5871},
  numpages = {0},
  year = {2000},
  month = {Jun},
  publisher = {American Physical Society},
  doi = {10.1103/PhysRevLett.84.5868},
  url = {https://link.aps.org/doi/10.1103/PhysRevLett.84.5868}
}

@article{Demokritov:2006bx,
author = {Demokritov, S O and Demidov, V E and Dzyapko, O and Melkov, G A and Serga, A A and Hillebrands, B and Slavin, A N},
title = {{Bose{\textendash}Einstein condensation of quasi-equilibrium magnons at room temperature under pumping}},
journal = {Nature},
year = {2006},
volume = {443},
number = {7110},
pages = {430--433},
doi = {10.1038/nature05117},
language = {English},
url = {http://www.nature.com/articles/nature05117},
uri = {\url{papers3://publication/doi/10.1038/nature05117}}
}

@article{Sonin_Adv_2010,
    author = {E.B. Sonin},
    title = {Spin currents and spin superfluidity},
    journal = {Advances in Physics},
    volume = {59},
    number = {3},
    pages = {181--255},
    year = {2010},
    publisher = {Taylor \& Francis},
    doi = {10.1080/00018731003739943},
}

@ARTICLE{Yuan_SciAdv_2018,
    title = {Experimental signatures of spin superfluid ground state in canted antiferromagnet {Cr$_2$O$_3$} via nonlocal spin transport},
   author       = " W. Yuan  and Q. Zhu  and T. Su  and Y. Yao  and W. Xing  and Y. Chen  and Y. Ma  and X. Lin  and J. Shi  and R. Shindou  and X. C. Xie  and W. Han ", 
   year         = "2018", 
   journal      = "Sci. Adv.", 
   volume       = "4", 
   pages        = "eaat1098",
}

@ARTICLE{Nikolaev_NatCom_2024,
    title = {Resonant generation of propagating second-harmonic spin waves in nano-waveguides},
   author       = " Nikolaev, K. O. and Lake, S. R. and Schmidt, G. and Demokritov, S. O. and Demidov, V. E.", 
   year         = "2024", 
   journal      = "Nat. Commun.", 
   volume       = "15", 
   pages        = "1827",
}

@ARTICLE{Rezende_IEEE_1990,
  author={Rezende, S.M. and de Aguiar, F.M.},
  journal={Proceedings of the IEEE}, 
  title={Spin-wave instabilities, auto-oscillations, and chaos in yttrium-iron-garnet}, 
  year={1990},
  volume={78},
  number={6},
  pages={893-908},
  }

@article{Anderson:1995dw,
author = {Anderson, M H and Ensher, J R and Matthews, M R and Wieman, C E and Cornell, E A},
title = {{Observation of Bose-Einstein Condensation in a Dilute Atomic Vapor}},
journal = {Science},
year = {1995},
volume = {269},
number = {5221},
pages = {198--201},
month = jul,
publisher = {American Association for the Advancement of Science},
doi = {10.1126/science.269.5221.198}}

@article{Suhl1957,
author = {Suhl, H.},
doi = {},
issn = {},
journal = {J. Phys. Chem. Solids},
month = {},
number = {},
pages = {209},
publisher = {American Physical Society},
title = {{The theory of ferromagnetic resonance at high signal powers}},

volume = {1},
year = {1957}
}

@article{Mathieu_PhysRevB_2003,
  title = {Brillouin light scattering analysis of three-magnon splitting processes in yttrium iron garnet films},
  author = {Mathieu, Christoph and Synogatch, Valeri T. and Patton, Carl E.},
  journal = {Phys. Rev. B},
  volume = {67},
  issue = {10},
  pages = {104402},
  numpages = {8},
  year = {2003},
  month = {Mar},
  publisher = {American Physical Society},
  doi = {10.1103/PhysRevB.67.104402},
  url = {https://link.aps.org/doi/10.1103/PhysRevB.67.104402}
}

@article{Romero_PhysRevB_2009,
  title = {Three-magnon splitting and confluence processes for spin-wave excitations in yttrium iron garnet films: Wave vector selective {Brillouin} light scattering measurements and analysis},
  author = {Ord\'o\~nez-Romero, C\'esar L. and Kalinikos, Boris A. and Krivosik, Pavol and Tong, Wei and Kabos, Pavel and Patton, Carl E.},
  journal = {Phys. Rev. B},
  volume = {79},
  issue = {14},
  pages = {144428},
  numpages = {8},
  year = {2009},
  month = {Apr},
  publisher = {American Physical Society},
  doi = {10.1103/PhysRevB.79.144428},
  url = {https://link.aps.org/doi/10.1103/PhysRevB.79.144428}
}

@article{Kurebayashi_NatMat_2011,
  title = {Controlled enhancement of spin-current emission by three-magnon splitting},
  author = {Kurebayashi, Hidekazu and Dzyapko, Oleksandr and Demidov, Vladislav E. and Fang, Dong and Ferguson, A. J. and Demokritov, Sergej O.},
  journal = {Nature Mater.},
  volume = {10},
  issue = {14},
  pages = {660},
  numpages = {},
  year = {2011},
  month = {},
}

@article{Schultheiss_PhysRevLett_2009,
  title = {Direct Current Control of Three Magnon Scattering Processes in Spin-Valve Nanocontacts},
  author = {Schultheiss, H. and Janssens, X. and van Kampen, M. and Ciubotaru, F. and Hermsdoerfer, S. J. and Obry, B. and Laraoui, A. and Serga, A. A. and Lagae, L. and Slavin, A. N. and Leven, B. and Hillebrands, B.},
  journal = {Phys. Rev. Lett.},
  volume = {103},
  issue = {15},
  pages = {157202},
  numpages = {4},
  year = {2009},
  month = {Oct},
  publisher = {American Physical Society},
  doi = {10.1103/PhysRevLett.103.157202},
  url = {https://link.aps.org/doi/10.1103/PhysRevLett.103.157202}
}

@article{Rao_PhysRevLett_2023,
  title = {Unveiling a Pump-Induced Magnon Mode via Its Strong Interaction with Walker Modes},
  author = {Rao, J. W. and Yao, Bimu and Wang, C. Y. and Zhang, C. and Yu, Tao and Lu, Wei},
  journal = {Phys. Rev. Lett.},
  volume = {130},
  issue = {4},
  pages = {046705},
  numpages = {7},
  year = {2023},
  month = {Jan},
  publisher = {American Physical Society},
  doi = {10.1103/PhysRevLett.130.046705},
  url = {https://link.aps.org/doi/10.1103/PhysRevLett.130.046705}
}

@article{Zhang_PhysRevApplied_2023,
  title = {Control of Magnon-Polariton Hybridization with a Microwave Pump},
  author = {Zhang, Chao and Rao, Jinwei and Wang, C.Y. and Chen, Z.J. and Zhao, K.X. and Yao, Bimu and Xu, Xu-Guang and Lu, Wei},
  journal = {Phys. Rev. Appl.},
  volume = {20},
  issue = {2},
  pages = {024074},
  numpages = {8},
  year = {2023},
  month = {Aug},
  publisher = {American Physical Society},
  doi = {10.1103/PhysRevApplied.20.024074},
  url = {https://link.aps.org/doi/10.1103/PhysRevApplied.20.024074}
}

@article{Wang_NatPhys_2024,
  title = {Enhancement of magnonic frequency combs by exceptional points},
  author = {Wang, C. and Rao, J. and Chen, Z. and others},
  journal = {Nat. Phys.},
  volume = {20},
  issue = {},
  pages = {1139-1144},
  numpages = {},
  year = {2024},
  month = {},
  publisher = {},
  url = {https://doi.org/10.1038/s41567-024-02478-0}
}

@article{RAMESHTI_Reports_2022,
title = {Cavity magnonics},
journal = {Physics Reports},
volume = {979},
pages = {1-61},
year = {2022},
note = {Cavity Magnonics},
issn = {0370-1573},
doi = {https://doi.org/10.1016/j.physrep.2022.06.001},
url = {https://www.sciencedirect.com/science/article/pii/S0370157322002460},
author = {Babak {Zare Rameshti} and Silvia {Viola Kusminskiy} and James A. Haigh and Koji Usami and Dany Lachance-Quirion and Yasunobu Nakamura and Can-Ming Hu and Hong X. Tang and Gerrit E.W. Bauer and Yaroslav M. Blanter}
}

@article{Li_PhysRevLett_2019,
  title = {Strong Coupling between Magnons and Microwave Photons in On-Chip Ferromagnet-Superconductor Thin-Film Devices},
  author = {Li, Yi and Polakovic, Tomas and Wang, Yong-Lei and Xu, Jing and Lendinez, Sergi and Zhang, Zhizhi and Ding, Junjia and Khaire, Trupti and Saglam, Hilal and Divan, Ralu and Pearson, John and Kwok, Wai-Kwong and Xiao, Zhili and Novosad, Valentine and Hoffmann, Axel and Zhang, Wei},
  journal = {Phys. Rev. Lett.},
  volume = {123},
  issue = {10},
  pages = {107701},
  numpages = {6},
  year = {2019},
  month = {Sep},
  publisher = {American Physical Society},
  doi = {10.1103/PhysRevLett.123.107701},
  url = {https://link.aps.org/doi/10.1103/PhysRevLett.123.107701}
}

@article{Shklovskij_PhysRevB_2018,
  title = {Nonlinear relaxation between magnons and phonons in insulating ferromagnets},
  author = {Shklovskij, Valerij A. and Mezinova, Viktoriia V. and Dobrovolskiy, Oleksandr V.},
  journal = {Phys. Rev. B},
  volume = {98},
  issue = {10},
  pages = {104405},
  numpages = {7},
  year = {2018},
  month = {Sep},
  publisher = {American Physical Society},
  doi = {10.1103/PhysRevB.98.104405},
  url = {https://link.aps.org/doi/10.1103/PhysRevB.98.104405}
}

@article{Wang_PhysRevLett_2021,
  title = {Magnonic Frequency Comb through Nonlinear Magnon-Skyrmion Scattering},
  author = {Wang, Zhenyu and Yuan, H. Y. and Cao, Yunshan and Li, Z.-X. and Duine, Rembert A. and Yan, Peng},
  journal = {Phys. Rev. Lett.},
  volume = {127},
  issue = {3},
  pages = {037202},
  numpages = {7},
  year = {2021},
  month = {Jul},
  publisher = {American Physical Society},
  doi = {10.1103/PhysRevLett.127.037202},
  url = {https://link.aps.org/doi/10.1103/PhysRevLett.127.037202}
}

@article{Li_JAP_2020,
    author = {Li, Yi and Zhang, Wei and Tyberkevych, Vasyl and Kwok, Wai-Kwong and Hoffmann, Axel and Novosad, Valentine},
    title = {Hybrid magnonics: Physics, circuits, and applications for coherent information processing},
    journal = {Journal of Applied Physics},
    volume = {128},
    number = {13},
    pages = {130902},
    year = {2020},
    month = {10},
    issn = {0021-8979},
    doi = {10.1063/5.0020277},

}

@article{Fletcher_PhysRev_1959,
  title = {Identification of the Magnetostatic Modes of Ferrimagnetic Resonant Spheres},
  author = {Fletcher, P. and Solt, I. H. and Bell, R.},
  journal = {Phys. Rev.},
  volume = {114},
  issue = {3},
  pages = {739--745},
  numpages = {0},
  year = {1959},
  month = {May},
  publisher = {American Physical Society},
  doi = {10.1103/PhysRev.114.739},
  url = {https://link.aps.org/doi/10.1103/PhysRev.114.739}
}

@article{Kaffash_QST_2023,
doi = {10.1088/2058-9565/ac9428},
url = {https://dx.doi.org/10.1088/2058-9565/ac9428},
year = {2022},
month = {oct},
publisher = {IOP Publishing},
volume = {8},
number = {1},
pages = {01LT02},
author = {Mojtaba T Kaffash and Dinesh Wagle and Anish Rai and Thomas Meyer and John Q Xiao and M Benjamin Jungfleisch},
title = {Direct probing of strong magnon–photon coupling in a planar geometry},
journal = {Quantum Science and Technology},
}

\end{document}